\documentclass[review]{elsarticle}

\usepackage{lineno,hyperref,graphicx,amsmath,amssymb,upgreek,booktabs,color,subfigure,siunitx}
\biboptions{numbers,sort&compress}
\journal{Computers \& Fluids}

\newcommand{\bi}{ \boldsymbol}

\newcommand{\nw}[1]{\textcolor{black}{#1}}
\newcommand{\rw}[1]{\textcolor{black}{#1}}

\begin{document}
\begin{frontmatter}

\title{Electro-vortex flow simulation using coupled meshes}

\author{Norbert Weber}
\author{Pascal Beckstein}
\author{Vladimir Galindo}
\author{Marco Starace}
\author{Tom Weier}

\address{Helmholtz-Zentrum Dresden - Rossendorf, Dresden, Germany}

\begin{abstract}
A numerical model for simulating electro-vortical flows in OpenFOAM is
developed. Electric potential and current are solved in coupled
solid-liquid conductors by a parent-child mesh technique. The magnetic
field is computed using a combination of Biot-Savart's law and
induction equation. Further, a PCG solver with special regularisation
for the electric potential is derived and implemented. Finally, a
performance analysis is presented and the solver is validated against
several test cases.
\end{abstract}


\end{frontmatter}


\section{Introduction}\label{intro}
Electro-vortex flow is highly relevant in many industrial
processes. Possible applications span from electromagnetic
stirring \cite{Ludwig2016} for grain size reduction in solidification
\cite{Raebiger2014, Franke2016} over electrode welding
\cite{Kazak2012}, electroslag welding, electroslag (re-)melting \cite{Kharicha2008,
  Kharicha2014}, vacuum arc melting \cite{Kharicha2015} to
electrolytic reduction (of e.g. aluminium \cite{Zhang2010}). Further,
many technical devices, as liquid fuses \cite{Cramer2004},
electric jet engines, arc furnaces \cite{Kazak2013}
and liquid metal batteries \cite{Weber2014b,Stefani2015,Weier2017} involve or
rely on electro-vortex flows. For an overview about such flows,
see \cite{Bojarevics1989,Davidson2001,Chudnovskii1989}.

Electro-vortex flow is not an instability. It develops at (or near) a
changing cross-section of a (liquid) conductor. Radial currents
produce, together with their own magnetic field, a Lorentz force,
which is non-conservative, i.e. its curl is not equal to zero. This force cannot be compensated totally by a pressure
gradient and therefore drives a flow. For an illustrative example,
see Shercliff \cite{Shercliff1970}.

Numerical simulation of electro-vortex flow is easy when modelling
only the fluid, or a non-conducting obstacle inside a fluid. However, in
most realistic cases, electric current passes from solid to liquid
conductors and vice versa. The electric potential in these regions
must therefore be solved in a coupled way. The classical, segregated
approach means solving an equation in each region, and coupling the
potential only at the interfaces by suitable boundary
conditions \cite{Weber2014b}. While that is easy to implement,
convergence is rather poor. An implicit coupling of the different regions by
block matrices is a sophisticated alternative for increasing
convergence \cite{Rusche2010}. However, it is memory-intensive and by
no means easy to implement.

In this article we will present an alternative effective option for region coupling in OpenFOAM.
We solve global variables (electric potential, current density) on a
global mesh with a variable electric conductivity according to the
underlying material. We then map the current density to the fluid
regions and compute the electromagnetic induced flow there. This parent-child mesh technique
was already used for the similar problem of thermal conduction
\cite{Beale2013,Beale2016} and just recently for the solution of
eddy-current problems with the finite volume method \cite{Beckstein2017}.

\section{Mathematical and numerical model}
\subsection{Overview}
The presented multi-region approach is based on a single phase
incompressible magnetohydrodynamic (MHD) model
\cite{Weber2013,Weber2014b}. The flow in the fluid is described by the
Navier-Stokes equation (NSE)
\begin{equation}\label{eqn:nse}
\frac{\partial \bi u}{\partial t} + \left( \bi u \cdot \nabla \right)
\bi u = - \nabla p + \nu \Delta \bi u + \frac{\bi J\times\bi B}{\rho},
\end{equation}
with $\bi u$ denoting the velocity, $t$ the time, $p$ the modified
pressure, $\nu$ the kinematic viscosity and $\rho$ the
density. \nw{The fluid flow is modelled as laminar only; adding a
  turbulence model is planned for the future.} We split the electric
potential $\phi$, the current density $\bi J$ and  the magnetic field
$\bi B$ into a constant (subscript 0) and induced part (lower case) as
\begin{align}
\phi &= \phi_0 + \varphi\\
\bi J &= \bi J_0 + \bi j\\
\bi B &= \bi B_0 + \bi b.
\end{align}
In order to determine the distribution of the constant part of the
electric potential $\phi_0$ we solve a Laplace
equation for the electric potential
\begin{equation}\label{eqn:potential0}
\nabla\cdot\sigma\nabla\phi_0 = 0
\end{equation}
on the global mesh. The above equation is obtained starting
from the Kirchhoff law of charge conservation ($\nabla\cdot \bi J_0 = 0$)
and $\bi J_0 = -\sigma\nabla\phi_0$.
Note that the conductivity $\sigma$ is a field and
not a constant, because the equation is solved on the full
geometry. \nw{During mesh generation, it is ensured that the border
between two materials always coincide with a face between
two neighbouring cells.} The global current density is then calculated
as
\begin{equation}\label{eqn:scurrent}
\bi J_0 = -\sigma\nabla\phi_0
\end{equation}
and mapped to the fluid region. Afterwards, the constant magnetic
field is determined as described in section \ref{ch:magField} only in
the fluid.

Often it is sufficient to calculate only the constant current and magnetic field.
Nevertheless, our solver also allows to compute their induced counterparts, e.g.
for simulating the Tayler instability \cite{Vandakurov1972,Tayler1973,Stefani2011,Seilmayer2012,Bonanno2012,Ruediger2013,Weber2014,Herreman2015,Stefani2018}.
The scheme is similar to that described above: in a first step, the induced electric potential
$\varphi$ is determined by solving a Poisson equation
\begin{equation}\label{eqn:potential}
\nabla\cdot\sigma\nabla\varphi = \nabla\cdot \sigma(\bi u\times\bi B)
\end{equation}
after mapping the source term $\bi u\times\bi B$ to the global mesh. The
induced current can be computed taking into account Ohm's law
\begin{equation}\label{eqn:icurrent}
\bi j = \sigma(-\nabla\varphi + \bi u\times\bi B).
\end{equation}
After mapping $\bi j$ to the fluid mesh we determine the induced
magnetic field as described in section \ref{ch:magField}.

\begin{figure}[t!]
\centering
  \includegraphics[width=0.9\textwidth]{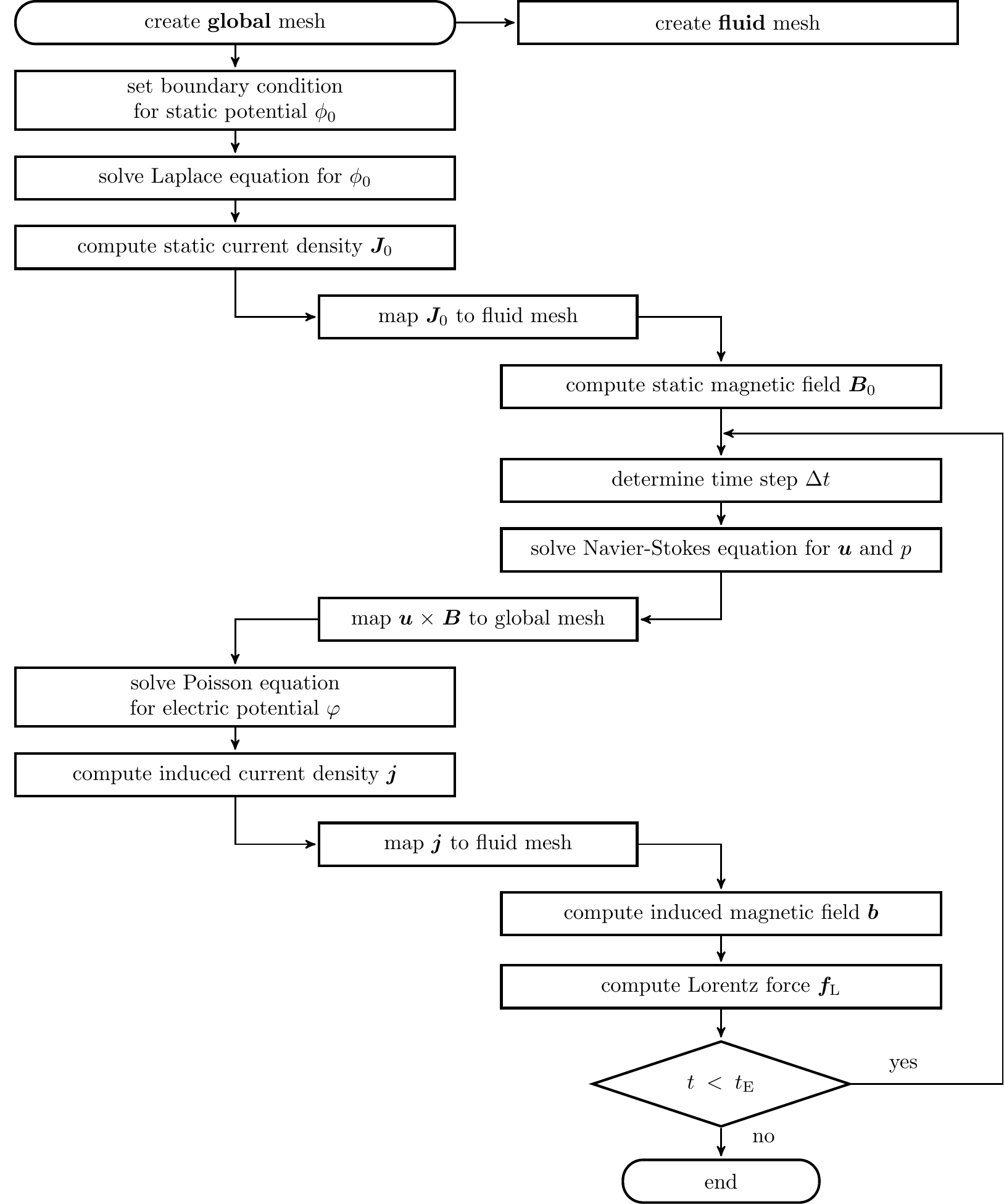}
\caption{Flowchart of the simulation model.}
\label{f:flowchart}
\end{figure}

Our model is not capable of describing AC currents, because we use the
quasi-static approximations by neglecting the temporal derivation of
the vector potential ($d\bi a/dt=0$) and magnetic field ($d\bi b /dt =
0$) \cite{Bandaru2016}. For a detailed flowchart of the model, please
refer to figure \ref{f:flowchart}.

\subsubsection{Computation of the magnetic field}\label{ch:magField}
For the computation of both, the constant part of the magnetic field
$\bi B_0$ and its induced counterpart $\bi b$ we use
the inversion of Amp\`ere's law, the Biot-Savart integral
\begin{equation}\label{eqn:biotsavart}
\bi B(\bi r) = \frac{\mu_0}{4\pi}\int \frac{\bi J(\bi
  r')\times(\bi r
  - \bi r')}{|\bi r - \bi r'|^3}dV'
\end{equation}
to determine both from the current density $\bi J$. This
integro-differential approach was proposed by Meir and
Schmidt \cite{Meir1994,Meir1996,Schmidt1999,
  Meir1999,Meir1999b,Meir2004} and later used for describing dynamos
\cite{Xu2004,Xu2004a,Xu2008} and the Tayler instability
\cite{Weber2013}.

In order to obtain the magnetic field in one single cell (at the position $\bi r$), the electric current densities
of all other cells (at the position $\bi r'$) have to be integrated. The number of
operations is therefore equal to the number of cells squared. This way of
computation is extremely costly. We will explain here
several ways for a speed up of the procedure. Solving Biot-Savart's integral on a
coarser grid, recalculating it every n$th$ time step, and an appropriate
parallelisation \cite{Weber2013} are the most simple ways.

\begin{figure}[tbh]
\centering
\subfigure[]{\includegraphics[height=2.4cm]{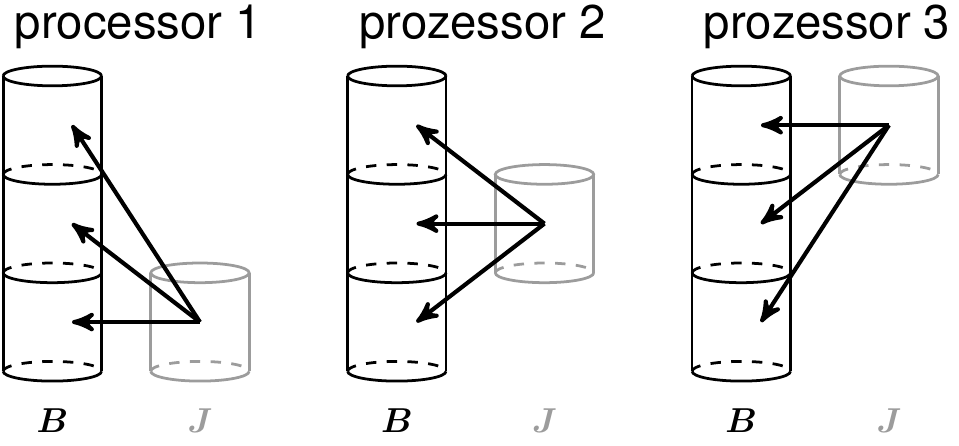}}\hfill
\subfigure[]{\includegraphics[height=2.4cm]{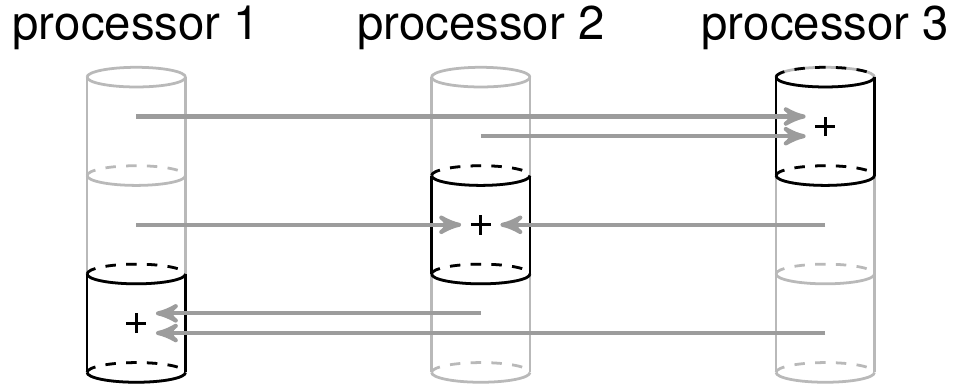}}
\caption{Each processor computes a full magnetic field from its local
current $\bi J$ (a), receives afterwards only its local $\bi B$ from
all other processors and adds it up (b).}
\label{f10}
\end{figure}

The parallelisation is implemented in OpenFOAM using MPI. Basically,
each processor contains only the current density of its \emph{local}
cells. With this, it computes the magnetic field for the \emph{full}
geometry (see figure \ref{f10}a). Finally, the field $\bi B$ of each
cell has to be summed up over all processors. This might be done
using the MPI function ALLREDUCE, resulting in a correct and global
$\bi B$ on all processors. However, this is not necessary, because a
single processor needs only its \emph{local} $\bi B$ for further
computation. Therefore, each processor receives only its \emph{local}
magnetic field from all other processors and adds up all contributions
given. The communication process is illustrated in figure \ref{f10}b.

Increasing the speed-up considerably is possible by computing
Biot-Savart's integral only on the boundaries and solving the
induction equations \cite{Weber2015b, Stefani2016}
\begin{align}
0&=\Delta\bi B_0\label{eqn:inductionB0}\\
0&= \frac{1}{\sigma\mu_0}\Delta\bi b + \nabla\times(\bi u\times\bi B_0) +
\nabla\times(\bi u\times\bi b)\label{eqn:inductionb}
\end{align}
for the constant and induced magnetic field in the quasi-static limit
\cite{Bandaru2016}.

An even faster alternative is shifting the problem from the magnetic
field $\bi B$ to the vector
potential $\bi A$ using the relation $\bi B =\nabla\times\bi A$.
Similar to Biot-Savart's law for $\bi B$, the vector potential can be
determined by Green's identity \cite{Santalo1993}:
\begin{equation}\label{eqn:biotA}
\bi A(\bi r) = \frac{\mu_0}{4\pi}\int\frac{\bi J(\bi r')}{|\bi r -\bi
  r'|}dV'.
\end{equation}
Please note that this formula is much cheaper to compute than
Biot-Savart's law (equation \ref{eqn:biotsavart})
\cite{Weber2017,Weber2017a}.

The transport equations for the vector potential are derived from
Amp\`ere's law, $\bi B=\nabla\times \bi A$, Ohm's law
\cite{Moreau1990}
and using the Coulomb gauge condition $\nabla\cdot\bi A=0$ as
\begin{align}
0&=\frac{1}{\sigma\mu_0}\Delta\bi A_0 -\nabla\phi_0\label{eqn:inductionEqnA0}\\
0&= \frac{1}{\sigma\mu_0}\Delta\bi a + \bi u\times \bi B_0 +
\bi u\times (\nabla\times\bi a) - \nabla\varphi.\label{eqn:inductionEqna}
\end{align}

\rw{
Basically all mentioned approaches of determining $\bi B$ based on the
equations (\ref{eqn:biotsavart}) till (\ref{eqn:inductionEqna}) are equal
from a physical point of view. But due to the way they are discretised and
numerically solved, there will be differences in both accuracy and calculation time.
While being the most expensive method, calculating the magnetic field by means of
Biot-Savart's law also gives the most accurate result. This stems from the fact that
the integral equation (\ref{eqn:biotsavart}) represents an exact solution for
$\bi B$ which is only numerically integrated for a finite number of cells. As already
mentioned, a computationally less expensive evaluation can be achieved with the help
of the magnetic vector potential $\bi A$ and Green's identity (\ref{eqn:biotA}),
where the complexity of the integrand is reduced compared to equation
(\ref{eqn:biotsavart}). Despite of equation (\ref{eqn:biotA}) also being an exact
solution, the subsequent calculation of $\bi B = \nabla \times \bi A$ introduces
an additional layer of discretisation errors from cell averaging and face interpolation.}

\rw{
As outlined above, Biot-Savart's law may be used also in combination with equations
(\ref{eqn:inductionB0}) and (\ref{eqn:inductionb}) or Green's identity (\ref{eqn:biotA})
combined with equations (\ref{eqn:inductionEqnA0}) and (\ref{eqn:inductionEqna}),
while only boundary values of $\bi B$ or $\bi A$ are evaluated using the exact
integral equations. Internal values are then recovered from solving the related
differential equations. This drastically improves computational efficiency at
the cost of some accuracy. However, from figure \ref{f:flowchart} one can comprehend
that it is sufficient to calculate $\bi B_0$ or $\bi A_0$ only once at the beginning
of a simulation, whereas $\bi b$ or $\bi a$ needs to be updated recurringly while
marching in time. The most promising way of determining the total magnetic field
$\bi B$ is thus to compute its static part $\bi B_0$ once and solely using
Biot-Savart's law with the current density $\bi J_0$ and the induced part $\bi b$
mediately from $\bi b = \nabla \times \bi a$, whereby the solution of the induced
magnetic vector potential $\bi a$ is in turn based on the transport equation
(\ref{eqn:inductionEqna}) for which Dirichlet boundary conditions are derived
from evaluating Green's identity with the current density $\bi j$. This approach
has been used for all following calculations.}

\rw{
In this way, $\bi B_0$ is most accurate and $\bi b$ is repeatedly calculated
with minimum computational effort. Another important advantage of this realisation
is that the solenoidal nature of $\bi B$ is implicitly satisfied, as $\bi B_0$
results from an exact solution in shape of Biot-Savart's law and $\bi b$ is
calculated from the definition of the induced vector potential $\bi a$ with
$\nabla\cdot\bi b = \nabla\cdot(\nabla\times\bi a) \equiv 0$. Numerically, Gauss's law
$\nabla\cdot\bi B = 0$ is of course only met approximatively due to discretisation
errors. With linear interpolation, the corresponding finite volume approximation is
second order accurate. Additionally, for small magnetic Reynolds numbers, which are
typical for most liquid metal MHD flows on laboratory scale, $\bi b$ is usually small
compared to $\bi B_0$.}

\rw{
As opposed to using $\bi b = \nabla \times \bi a$, a direct solution of the
induction equation (\ref{eqn:inductionb}) for $\bi b$ would in general require
additional steps to ensure its solenoidal property. This is particularly true in
ideal or plasma MHD for time-dependent problems at high magnetic Reynolds numbers,
where the induction equation is dominated by convection \cite{Brackbill1980}. For
such cases it is usually necessary to adopt a special correction. An overview of
possible divergence cleaning methods can be found in \cite{Miyoshi2011}. The OpenFOAM
standard solver mhdFoam for example uses the projection method, which is well known
from the pressure-velocity coupling of the Navier-Stokes equations.}

\rw{
Accordingly, also the solution of equation (\ref{eqn:inductionEqna}) would generally
require some correction to maintain the solenoidal property of the magnetic vector
potential $\bi a$. However $\nabla\cdot\bi b = 0$ is a physical requirement, whereas
$\nabla\cdot\bi a = 0$ is just a gauge. If the Helmholtz decomposition
$\bi a = \nabla\times\bi\Pi + \nabla\Psi$ is consulted, it can be easily seen that
$\nabla\cdot\bi a = \nabla\cdot\nabla\Psi$ does not influence the magnetic field
$\bi b = \nabla\times\bi a = \nabla\times\nabla\times\bi\Pi$. Thus, even a weakly
satisfied Coulomb gauge $\nabla\cdot\bi a \approx 0$ would suffice to correctly
represent $\bi b$. Fixing the gauge is merely important to achieve a unique solution
for $\bi a$. Moreover, for diffusively dominated cases the Coulomb gauge may be
incorporated directly into Amp\`ere's law according to
$\nabla\times\nabla\times\bi a - \nabla(\nabla\cdot\bi a) = - \nabla\cdot\nabla\bi a = \mu_0\bi j$
as long as charge conservation $\nabla\cdot\bi j = 0$ is satisfied \cite{BiroPreis1989}.
This requirement is met with the solution of \ref{eqn:potential}.
In equation (\ref{eqn:inductionEqna}), the sum
$\bi u\times \bi B_0 + \bi u\times (\nabla\times\bi a) - \nabla\varphi$ corresponds
to the induced current density $\bi j/\sigma$ (\ref{eqn:icurrent}). If we explicitly
discretise these terms, they can be regarded as one source term for a Poisson equation,
whose system matrix is symmetric. For cases like this, it was demonstrated in
\cite{Beckstein2017} that indeed no additional divergence cleaning for $\bi a$ is required.}

\section{Discretisation}
Special attention must be paid to the discretisation of the Laplace
term $\nabla\cdot(\sigma\nabla\phi)$ of equation (\ref{eqn:potential0}) and
(\ref{eqn:potential}) because of the sharp jump in conductivity between
different materials. \nw{This jump is not smeared, but exactly
  reproduced in our model.} A linear interpolation of $\sigma$ would lead to
a wrong potential near the interface.

For a consistent application of the Gauss theorem to discretise the
equations \nw{(see \cite{openfoam2012})}, the electric conductivity is
interpolated harmonically. Knowing that the potential $\phi_f$ and the
normal current $(\bi j\cdot \bi n)_f$ must be continous from a cell
$P$ to its neighbour $N$, we find the conductivity at the face $f$ to be
\begin{equation}
\sigma_f =
\left(\dfrac{(\delta_P/\delta)}{\sigma_P}+\dfrac{(\delta_N/\delta)}{\sigma_N}\right)^{-1}
\end{equation}
with $\delta_i$ denoting the distance cell centre - face and $\delta$
the distance between both cell centres. In the quasi-static limit, this
exactly matches the embedded discretisation scheme which was derived in
\cite{Beckstein2017} to get a proper discretisation of the
Laplacian. \nw{For a more detailed discussion and similar
  discretisation of the thermal conductivity, see
\cite{Carson2005,Kumar2014,Marschall2012,Nabil2016}.}

Secondly, care must be taken when computing the gradient of the
potential to determine the current density as $\bi
J=-\sigma\nabla\phi$. In order to be able to use the Gauss theorem for discretisation, the
electric potential on the faces must be determined. Using the same
assumptions as for the harmonic interpolation described above, we identify the
electric potential at the face as
\begin{equation}
\phi_f = w\phi_P + (1-w)\phi_N
\end{equation}
with the interpolation weight
\begin{equation}
w = \frac{\delta_N\sigma_P}{\delta_P\sigma_N+\delta_N\sigma_P}.
\end{equation}
As before, this interpolation scheme corresponds to the embedded
discretisation of the gradient from \cite{Beckstein2017} in case of
the quasi-static assumption. All other discretisation schemes do not
need special attention. \rw{We use backward differencing for time
  discretisation, a mixed linear-upwind scheme for the convective term
  in equation (\ref{eqn:nse}) and second order linear discretisation for
  all other schemes.}

\section{Equation solvers}
The solution procedure of our model is illustrated in figure \ref{f:flowchart}.
As the Navier-Stokes equation is discretised and solved by means of the
PISO-algorithm \cite{Issa1985}, \rw{four} different Poisson equations need to be
addressed. This comprises the Laplace equation for the
static potential $\phi_0$, \rw{one Poisson equation for the vector potential $\bi a$,}
one Poisson equation for the potential $\varphi$ and
another Poisson equation for the fluid pressure $p$. Especially the latter two
are most commonly solved for Neumann boundary conditions. To improve the
overall robustness of the solution process in connection with the employed
parent-child mesh approach, we have implemented an alternative regularisation
technique for the iterative equation solvers in OpenFOAM, which is briefly
explained in the following.

\newcommand{\mat}[1]{\boldsymbol{\mathrm{#1}}}
\newcommand{\M}{\mat{M}}
\newcommand{\X}{\mat{\uppsi}}
\newcommand{\R}{\mat{r}}
\newcommand{\0}{\mat{0}}
\newcommand{\1}{\mat{1}}
\newcommand{\I}{\mat{I}}
\newcommand{\LO}{\lambda_1}
\newcommand{\Lk}{\lambda_k}
\newcommand{\VO}{\mat{v}_1}
\newcommand{\Vk}{\mat{v}_k}

The discretisation of a Poisson equation leads to a linear equation system
\begin{equation}\label{eqn:linsys}
\M\X = \R,
\end{equation}
where $\M \in \mathbb{R}^{n \times n}$ is a symmetric positive semi-definite
matrix, $\X \in \mathbb{R}^n$ is the discrete solution vector for either
$\varphi$ or $p$, and the right-hand side $\R \in \mathbb{R}^n$ mainly
represents the inhomogeneous part. Each row of the system (\ref{eqn:linsys}) is
related to one of $n$ cells. In case of a Neumann problem, the system matrix
will be singular and the solution is only defined up to an additive constant
vector. More specifically, the one-vector $\1 = (1,1, \dots, 1,1)^T$ lies in
the null space of the linear map $\M\X$. In other words, $\VO = \1/\sqrt{n}$ is
a normalized eigenvector corresponding to the eigenvalue $\LO = 0$ in
accordance with the identity $(\M - \LO\I)\VO = \0$.

\newcommand{\mati}[1]{\mathrm{#1}}
\newcommand{\mP}{\mati{m}_P}
\newcommand{\xP}{\mati{\uppsi}_P}
\newcommand{\xR}{\mati{\uppsi}_R}
\newcommand{\cR}{\mati{c}_R}
\newcommand{\rP}{\mati{r}_P}

In OpenFOAM such a singular matrix $\M$ is regularised by means of adding the
equation
\begin{equation}\label{eqn:refval}
\cR\xP = \cR\xR
\end{equation}
to the row which belongs to cell $P$, where $\cR$ is initially an arbitrary
coefficient, $\xP$ is the unknown solution and $\xR$ is a reference solution
for that cell. In order to slightly increase diagonal dominance of $\M$, $\cR$
is usually set to the diagonal coefficient of the matrix before adding the
equation: $\cR = \mP$. By specifying the reference value $\xR$, the solution
gets locally constrained in a weak sense. This approach is however extremely
sensitive to the smallest errors in the corresponding compatibility condition
of the Neumann problem. Such numerical errors may arise from the data exchange
between child and parent mesh due to interpolation.

A much more robust regularisation can be achieved by inverting the idea of the
so called Hotelling deflation \cite{Wilkinson1965}, which is actually a simple
technique to solve eigenproblems by selectively shifting single known
eigenvalues of a matrix to zero. Conversely, we may use the same procedure to
shift them also from zero to an arbitrary value, thus inflating the matrix.

\newcommand{\Mnew}{\widetilde{\mat{M}}}
\newcommand{\Lnew}{\widetilde{\lambda}_1}
\newcommand{\xk}{\mati{\uppsi}_k}

According to the spectral theorem for symmetric matrices \cite{Parlett1998}, it
is possible to decompose $\M$ based on its eigenvalues $\Lk$ and orthonormal
eigenvectors $\Vk$:
\begin{equation}\label{eqn:spectraldecomp}
\M = \sum_{k=1}^{n} \Lk\Vk\Vk^T = \LO\VO\VO^T + \sum_{k=2}^{n} \Lk\Vk\Vk^T.
\end{equation}
Using this decomposition we may then create a non-singular matrix $\Mnew$ using
only $\VO$ from above:
\begin{equation}\label{eqn:newm}
\Mnew = \M + \Lnew\VO\VO^T = \M + \Lnew\frac{1}{n}\1\1^T,
\end{equation}
where $\Lnew$ is any non-zero eigenvalue replacing $\LO$. It is important to
note that $\Mnew$ does not preserve the original sparsity pattern of $\M$,
which is usually undesired. Hence, a direct manipulation would not only mean a
waste of memory, but also a contraction in terms of the face addressing of
OpenFOAM. However, we may include the modification indirectly when computing
the matrix-vector product:
\begin{equation}\label{eqn:newmvec}
\Mnew\X = \M\X + \Lnew\frac{1}{n}\1\1^T\X = \M\X + \Lnew\frac{1}{n}\sum_{k=1}^{n}\xk\1,
\end{equation}
which is essentially the kernel of any iterative equation solver
\cite{Saad2011}. Furthermore parallelisation is straight-forward as the
exchange of the rightmost sum does only require little communication.

Taking the properties of $\M$ into consideration, it can be shown that all of
its eigenvalues are smaller or equal to twice the maximum of its diagonal
coefficients. Therefore we use the diagonal mean as modified eigenvalue
$\Lnew = \left<\mP\right>$, thus preserving the spectral radius of $\M$. Tests
with the preconditioned CG-method \cite{Saad2011} showed that the smoothness of
the numerical solution is preserved even if errors in the compatibility
condition exist. \nw{It is exactly this preservation of smoothness which
distinguishes our method from the original regularisation technique in OpenFOAM,
and which makes our method superior.}

\section{Results}
\subsection{Test case 1: speed-up of Biot-Savart's law}
In this section we present a performance analysis of the magnetic
field computation in a cylindrical geometry with an imposed current density
$\bi J$ (the other parts of the solver are switched off). The speedup
and scaling analysis is carried out on a cluster with Intel
8-Core Xeon 3,3 GHz CPUs cross linked with  \SI{40}{Gbit/s}
Infiniband. The solvers are compiled with OpenFOAM  2.2.0 and MPI 1.6.3.

\begin{figure}[tbh]
\centering
\subfigure{\includegraphics[height=4.2cm]{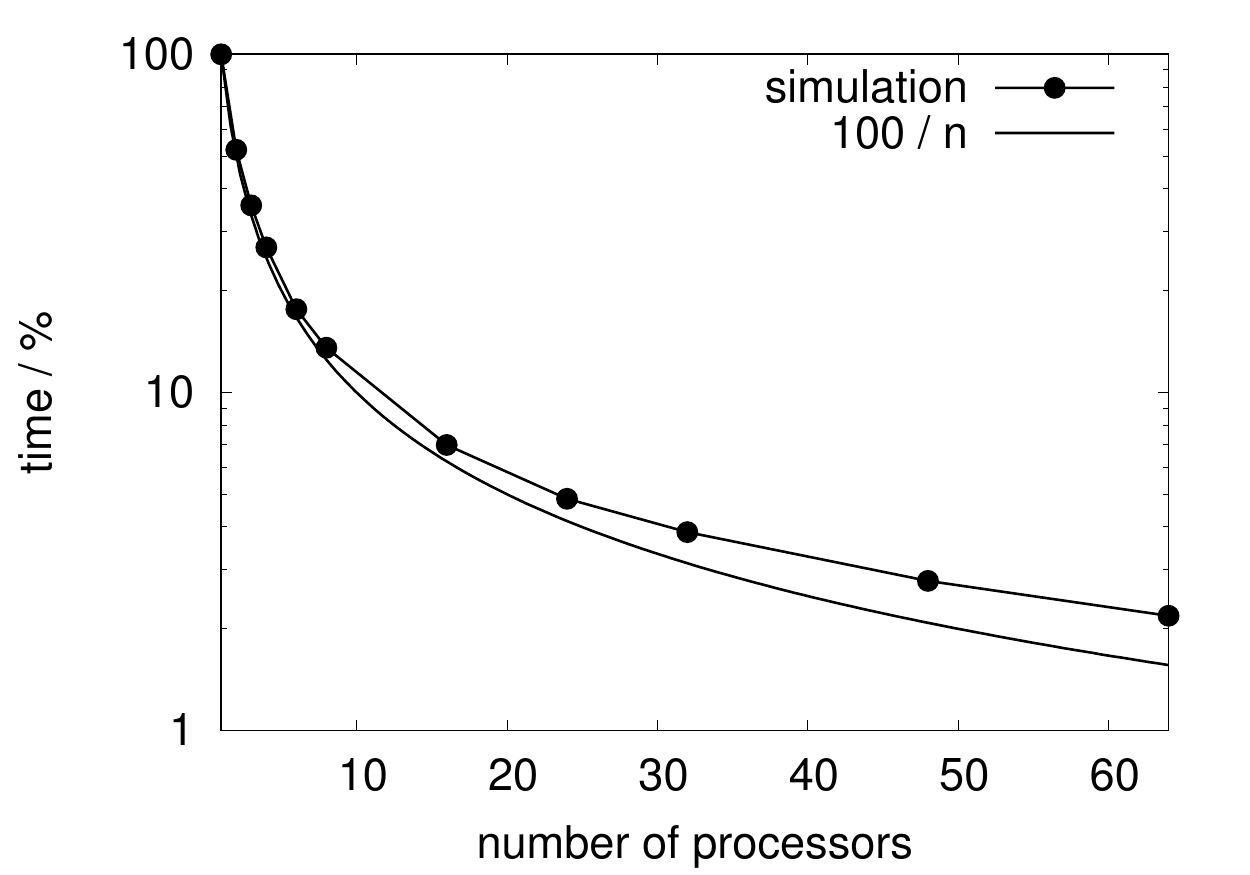}}\hfill
\subfigure{\includegraphics[height=4.2cm]{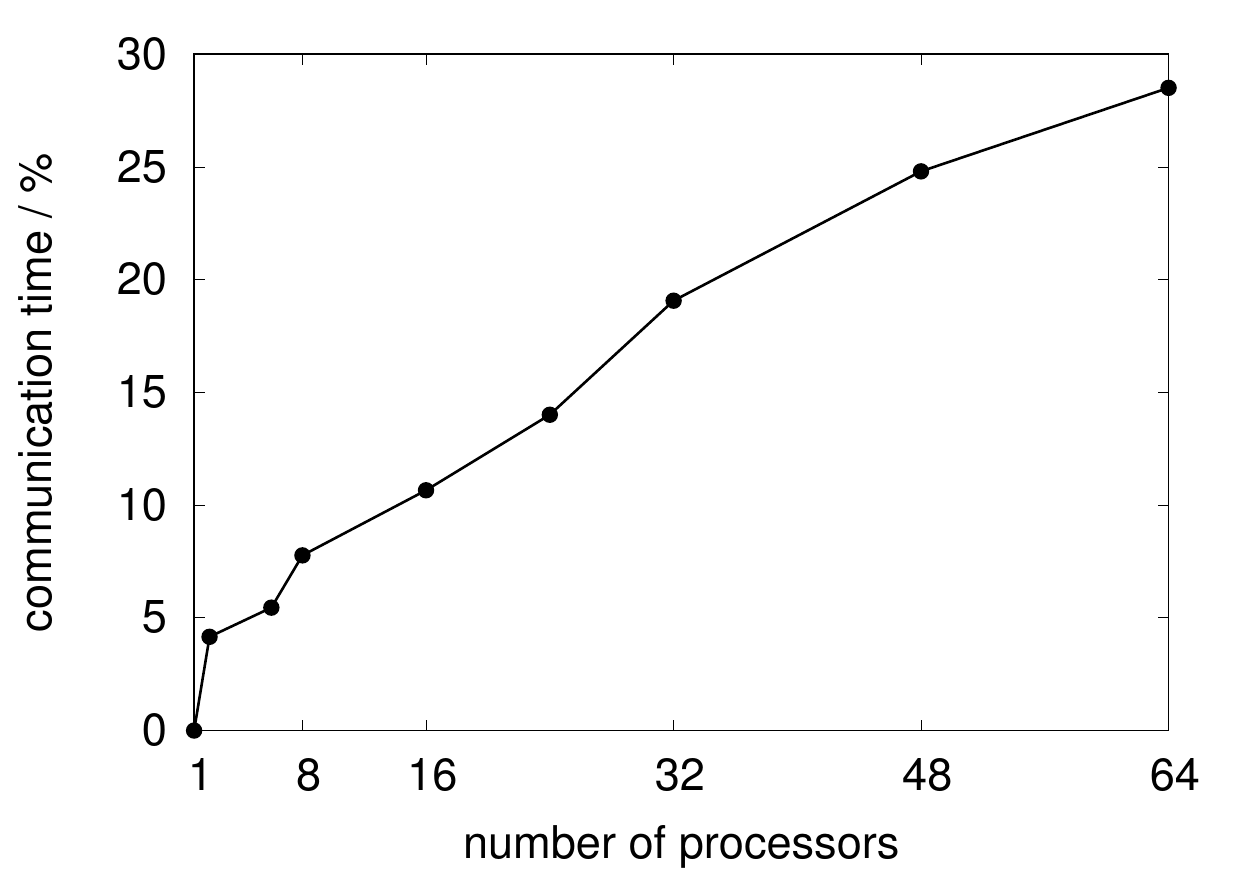}}
\caption{Computation time of Biot-Savart's law on 1 to 64 processors
  (a) and communication time divided by total time (b).}\label{f:speed1}
\end{figure}

In a first step we solve only Biot-Savart's law (equation
\ref{eqn:biotsavart}) for all cells and boundary faces -- on a changing
number of processors. The test case contains \num{352000}
cells. Figure \ref{f:speed1}a shows a good scaling up to
64 processors. The communication time is \SI{28}{\%} when using all 64
processors. In that case a single processor contains only \num{5\,500}
cells.

\begin{figure}[tbh]
\centering{}\includegraphics[height=5cm]{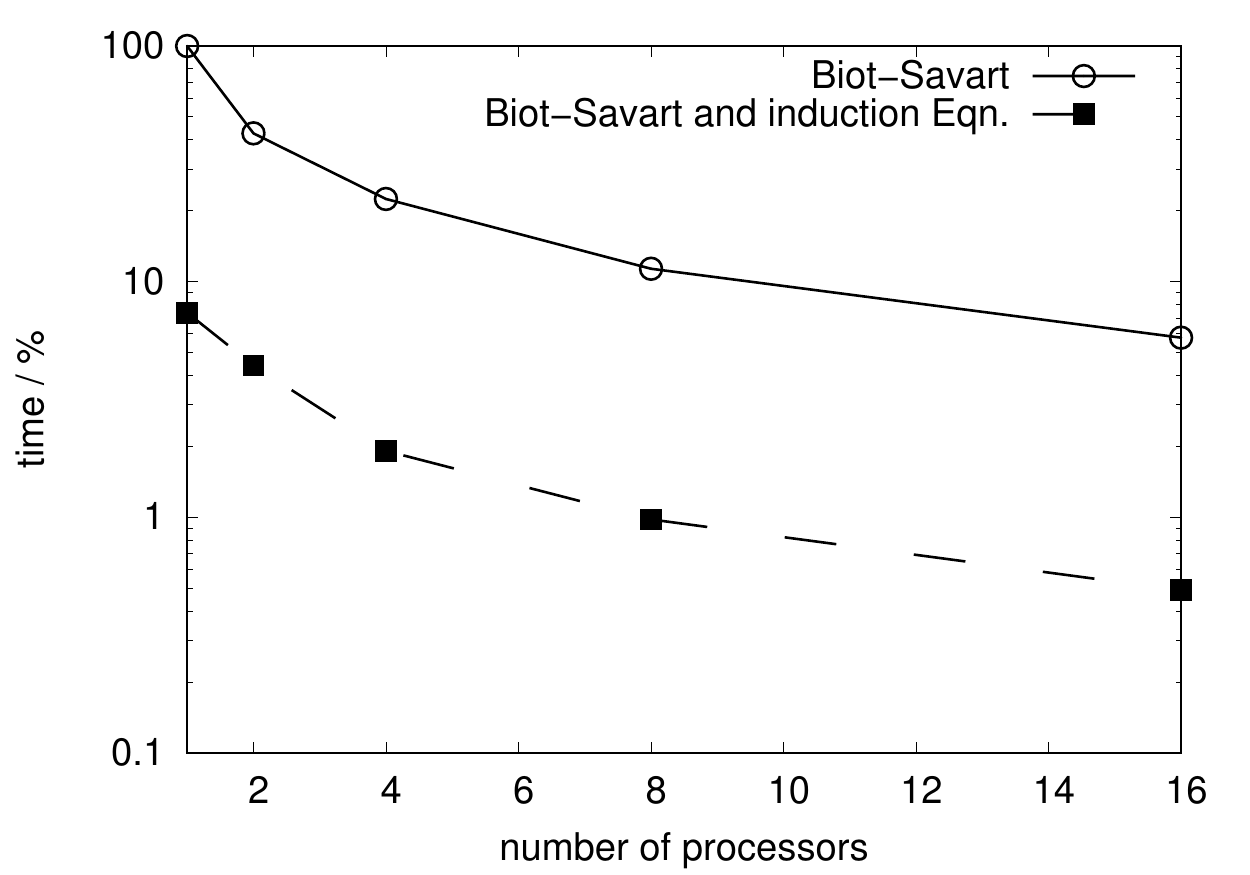}
\caption{Relative calculation time for the volume based Biot-Savart
  and
the surface based Biot-Savart combined with solving the induction
equation.}
\label{f:speed2}
\end{figure}

In a second test case, we use the same configuration again and compare
the full Biot-Savart integral with the method of solving the induction
equation (\ref{eqn:inductionB0}). For the latter, we compute Biot-Savart's
law only on the patches in order to obtain the correct boundary conditions.
Figure \ref{f:speed2} shows the relative computation
times (total cpu time/(cpu time for simulation in one processor)$\cdot$100\%)
for one to 16 processors. The method of using the Biot-Savart law on the
boundary regions only together with the solution of the corresponding induction
equation in the inner region scales very well, too; it is approximately
\num{13.5} times faster than the volume Biot-Savart method. Note that this
factor will probably increase for larger problems with more cells.

In a third case we use a mesh with \num{63200} cells and compare the
magnetic field with the vector potential approach. In both cases we
firstly compute the boundary conditions and solve then a transport
equation for $\bi A$ or $\bi B$ on a single processor 50 times. The
fastest result we obtain by using Biot-Savart for the
vector potential (equation \ref{eqn:biotA} and
\ref{eqn:inductionEqnA0}). Computing the magnetic field on the boundary
and solving the induction equation (\ref{eqn:biotsavart} and
\ref{eqn:inductionB0}) is five times slower. The volume-based
Biot-Savart is 84 times slower. Of course this holds only for the
Biot-Savart calculation; the differences for the whole solver, where
the flow simulation is included, will be smaller.

\subsection{Test case 2: current distribution in 2D}
\begin{figure}[h!]
\centering
 \subfigure[]{\includegraphics[width=0.15\textwidth]{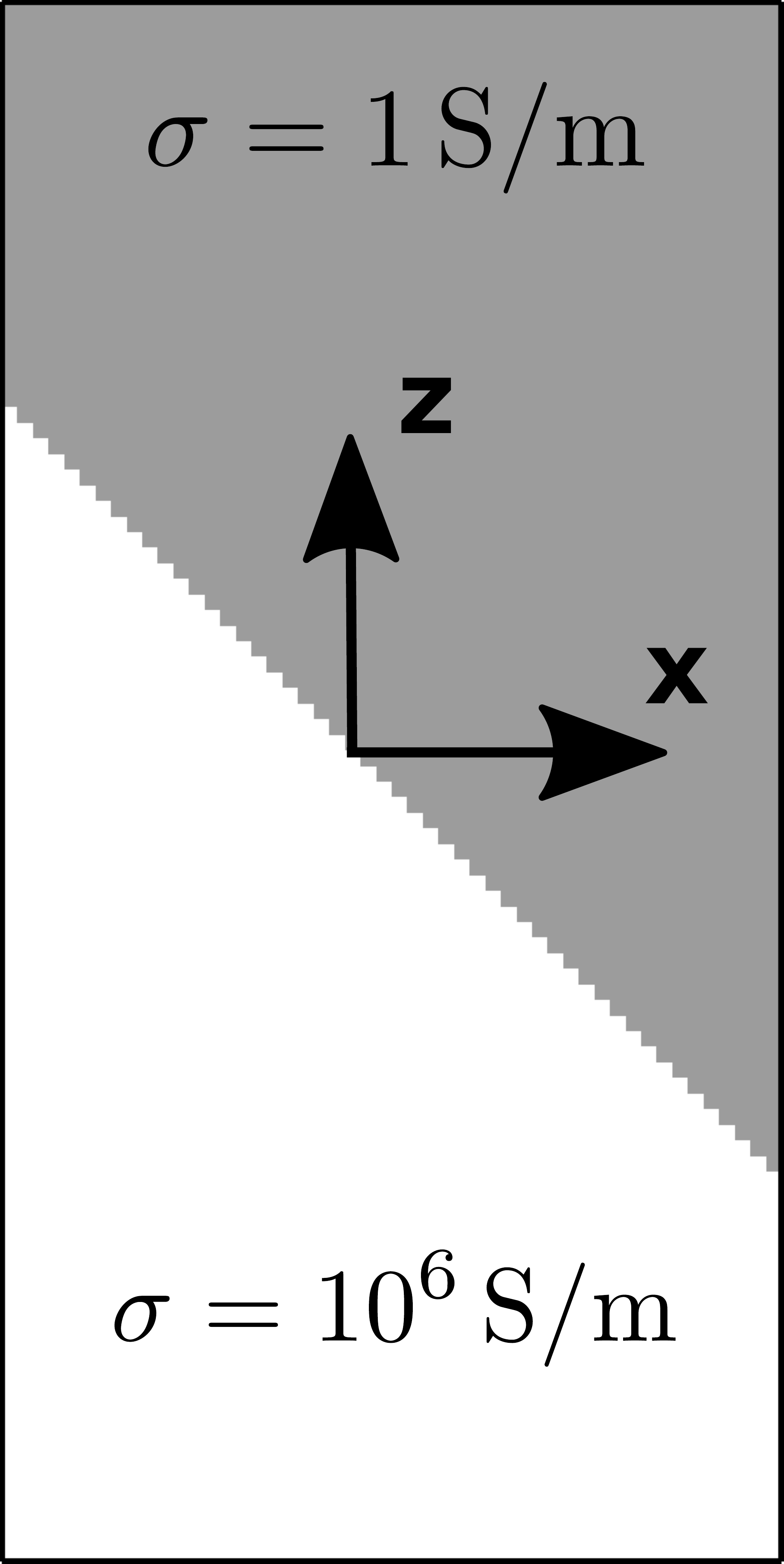}}\hfill
 \subfigure[]{\includegraphics[width=0.15\textwidth]{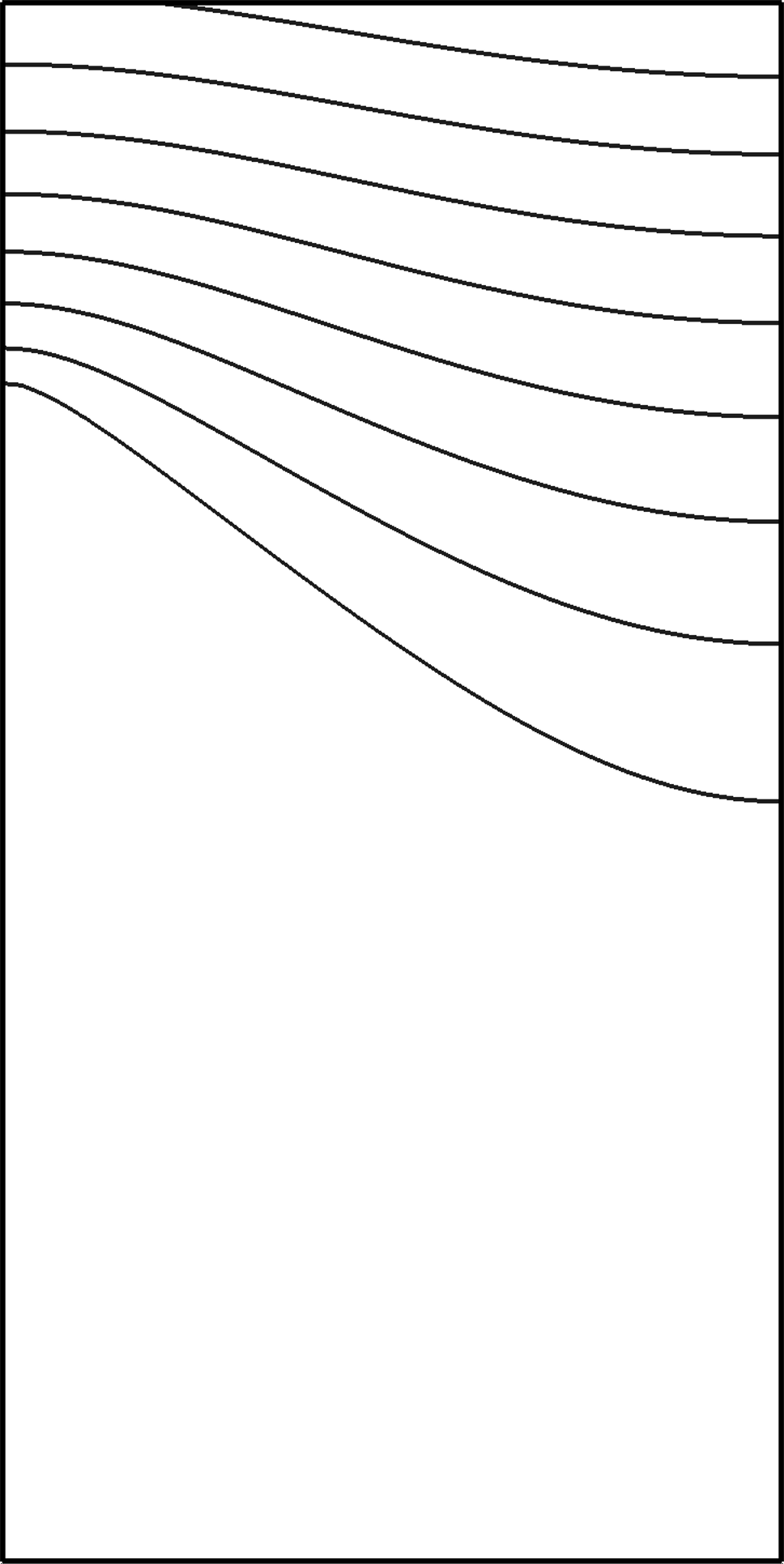}}\hfill
 \subfigure[]{\includegraphics[width=0.15\textwidth]{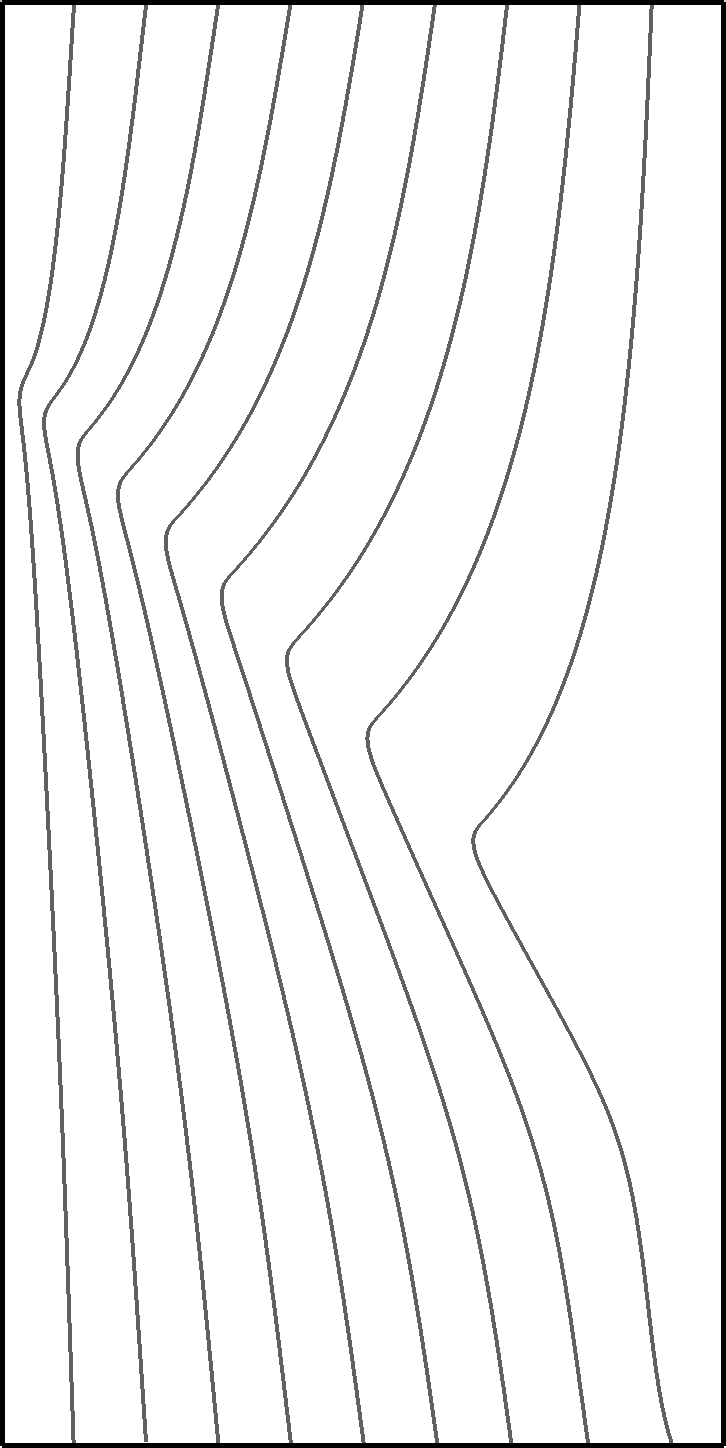}}\hfill
 \subfigure[]{\includegraphics[width=0.15\textwidth]{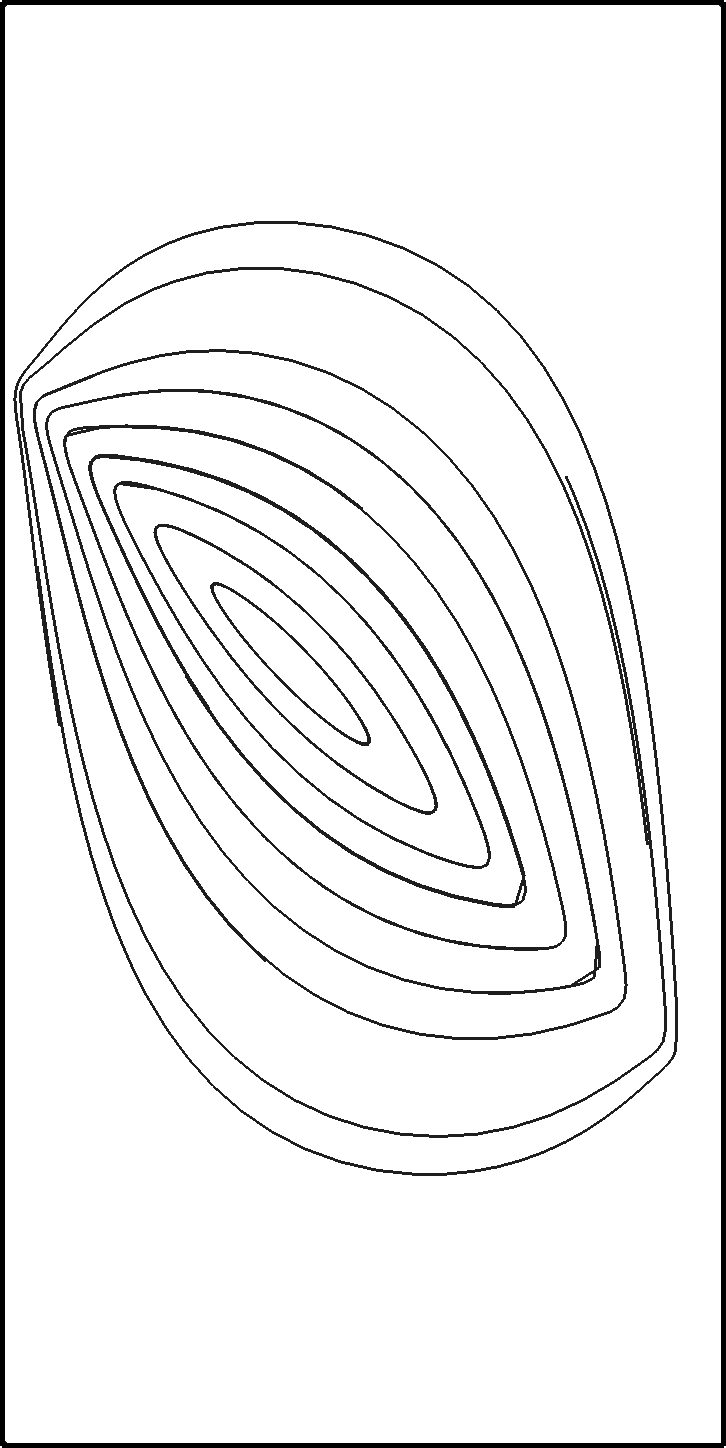}}
\caption{Conductivity distribution (a), electric potential (b), \nw{complete current density}
  (c) and \nw{disturbed current density} (d). The applied electrical current of \SI{1}{A}
  if flowing upwards.}
\label{f:2dcurrent}
\end{figure}

\begin{figure}[h!]
\centering
 \subfigure[]{\includegraphics[height=0.28\textwidth]{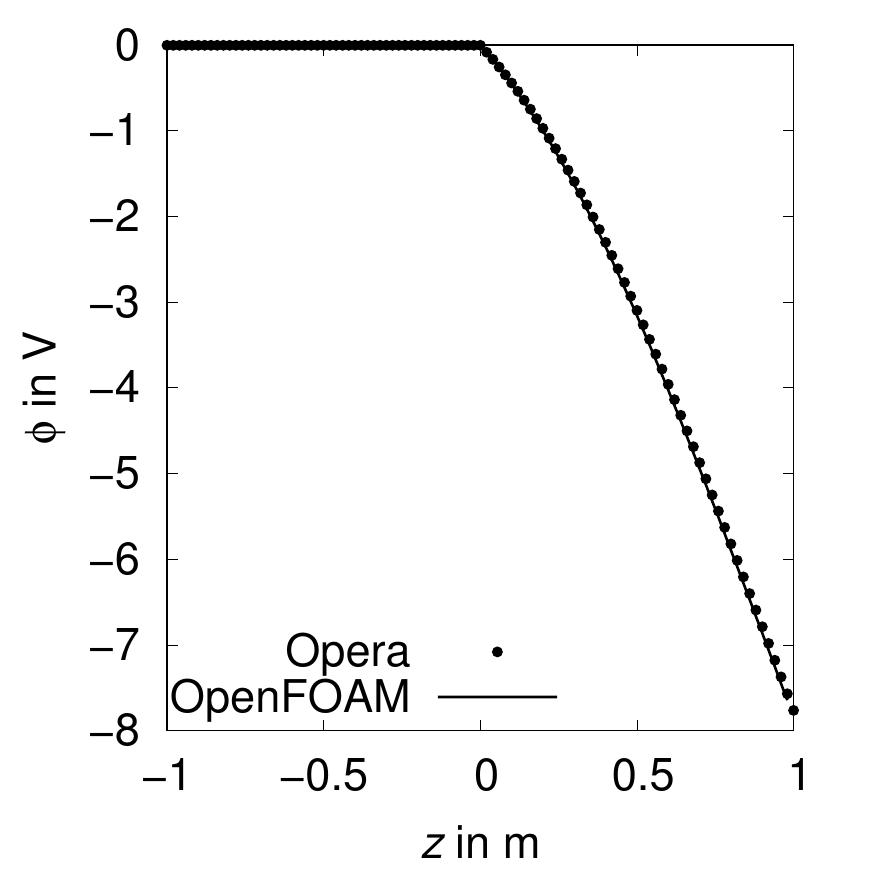}}\hfill
 \subfigure[]{\includegraphics[height=0.28\textwidth]{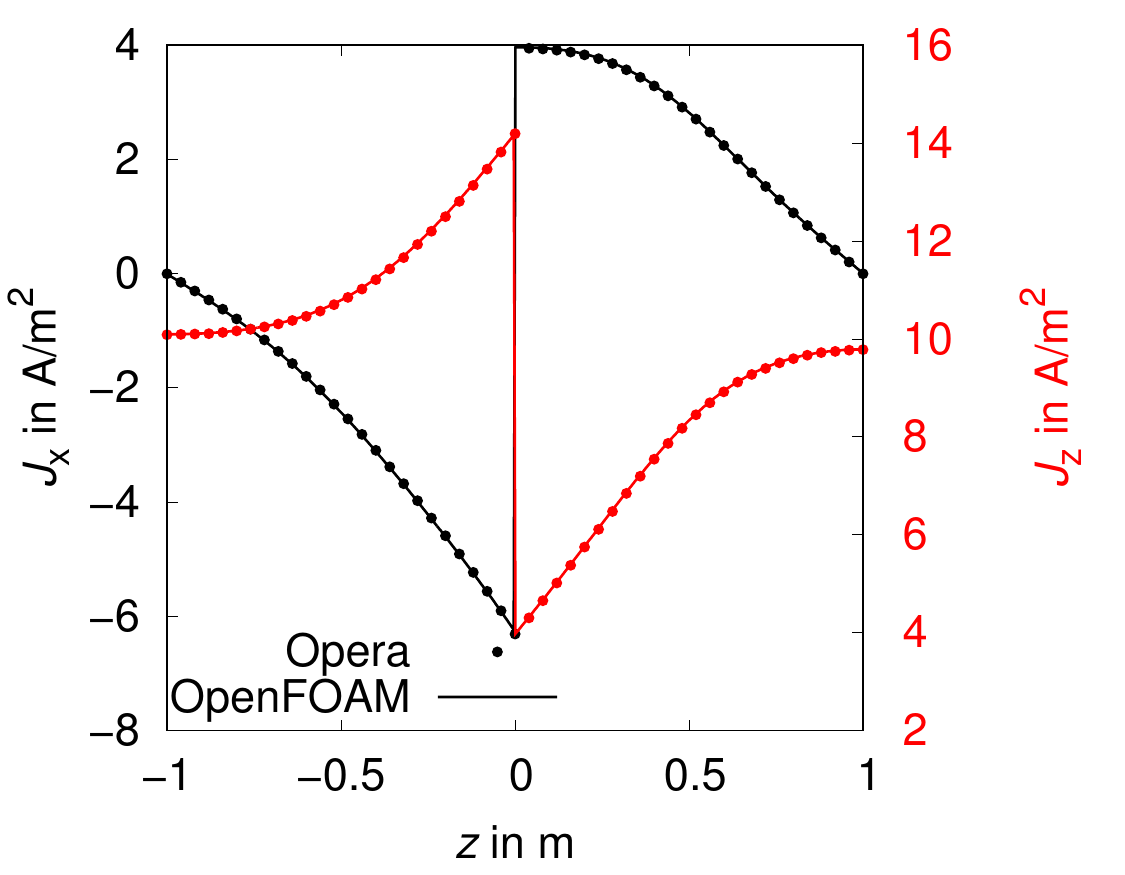}}\hfill
 \subfigure[]{\includegraphics[height=0.28\textwidth]{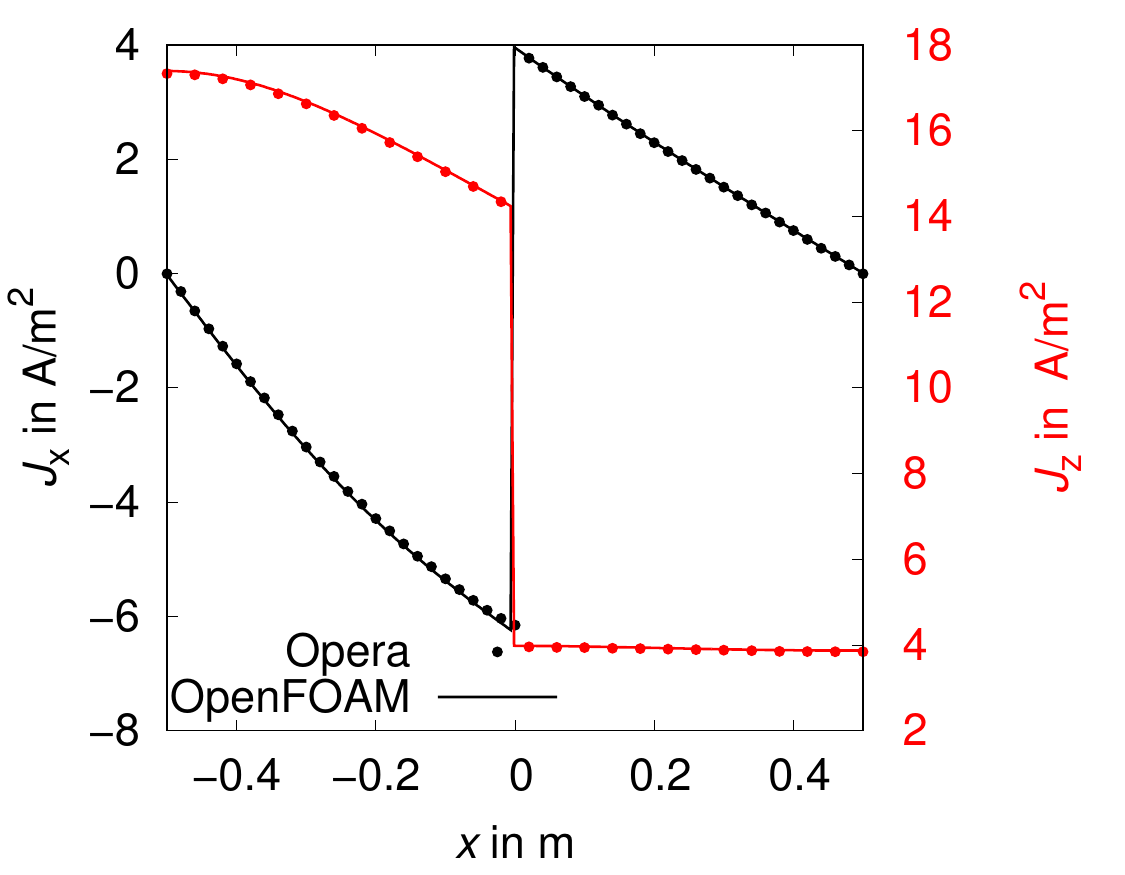}}
\caption{Electric potential (a) and horizontal current (b) along a
  centered vertical line for an applied current of \SI{1}{A} \nw{and
current density along a horizontal line at $z=0$.}}
\label{f:2dprofile}
\end{figure}

In a second test case the discretisation schemes for electric
conductivity and potential are validated by comparison with the
commercial
software Opera. We simulate a simple two-dimensional geometry
($1\times 2\times \SI{0.1}{m}$), consisting of two conductors of very
different conductivity with an inclined surface (inclination
\SI{45}{\degree}) -- see figure \ref{f:2dcurrent}a. A vertical current
of \SI{1}{A} is applied. Figure \ref{f:2dcurrent}b shows the
equipotential lines, figure \ref{f:2dcurrent}c the current lines and
\ref{f:2dcurrent}d the disturbed current. As expected, the current
lines concentrate in the area of high conductivity.
Figure \ref{f:2dprofile} shows the electric potential and the currents
along a vertical and horizontal
 centred line. The result of OpenFOAM and
Opera match very well. Obviously, Opera uses Dirichlet boundary
conditions for the electric potential (i.e. an equipotential
surface) -- so the same was done in OpenFOAM.

\subsection{Test case 3: electro-vortex flow in a cylindrical geometry}
Several model experiments \cite{Woods1971,Dementev1988,Dementev1992}
and similar analytical solutions \cite{Butsenieks1976,Millere1980} of
electro-vortex flow are known from  literature with most of them unfortunately
lacking detailed information. Here we will study the well reviewed
example of a thin electrode touching a cylindrical bath of liquid
metal \cite{Moshnyaga1980,Volokhonskii1991,Chudnovskii1989a}.
\begin{figure}[b!]
\centering
 \includegraphics[width=\textwidth]{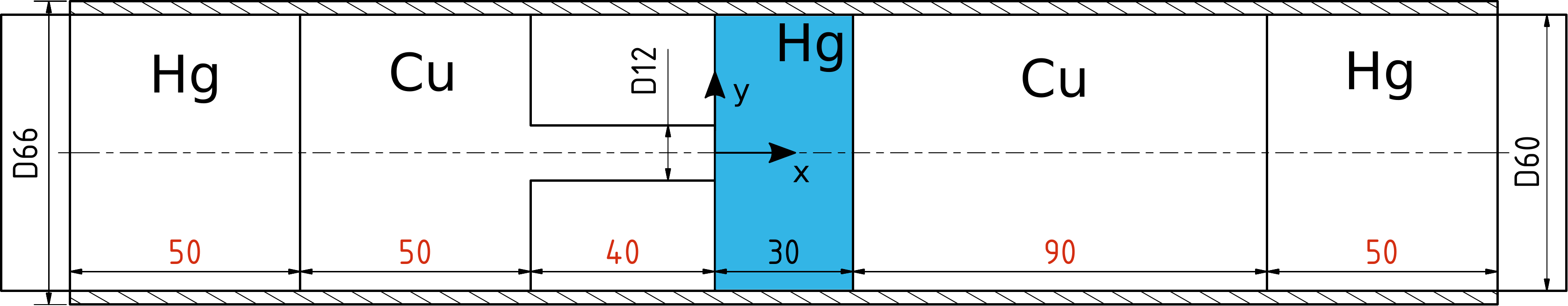}
\caption{Sketch of the experiment of Zhilin et al. \cite{Zhilin1986}.
The experiment is modelled with thick lateral current collectors which
are 3\,m long (but not shown in the image). The red dimensions (in mm) are estimated. The working section
(blue) with the symmetry axis $x$ is filled with liquid mercury.}
\label{f:zhilin_setup}
\end{figure}
 The
experiment was conducted at the Institute of Physics in Riga and
published by Zhilin et al. \cite{Zhilin1986}. Figure
\ref{f:zhilin_setup} illustrates the setup: a horizontal current passes
through a cylindrical bath of liquid mercury (colored in blue). One
copper electrode covers the whole surface, the other is reduced to a
small rod. The whole experiment is embedded into a steel pipe; two
mercury filled ``buffer zones provide for a smooth current
transition'' between external wires and the experiment. The axial
velocity along the cylinder axis is measured with a spacing of 1\,mm
in $x$-direction at $y=0$. The current is increased to up to
  1\,500\,A.

Unfortunately, the article does not provide any details about the
external current leads. They are therefore assumed to be infinitely
long. The measurements colored in red (fig. \ref{f:zhilin_setup}) were
not quoted by Zhilin et al. \cite{Zhilin1986}, but
estimated from the sketch. Similarly, the material properties were not
given;
they may vary considerably depending on the exact material/alloy.
We assume the copper conductivity to be $\sigma_\text{Cu}=58.5\cdot
10^6$\,S/m, the conductivity of mercury as $\sigma_\text{Hg}=1.04\cdot
10^6$\,S/m, its density as $\rho_\text{Hg}=13\,534$\,kg/m$^3$ \nw{and its
kinematic viscosity as $\nu=1.2\cdot 10^{-7}$\,m$^2$/s \cite{IAEA2008,Bettin2004}.} The
tube is made of ``stainless steel''; we assume therefore an electric
conductivity of $\sigma_\text{St} = 1.4\cdot10^{6}$\,S/m which is
typical for X5CrNi18-10. The tube works as potential divider --
only a part of the current passes through the mercury/copper.
\begin{figure}[t!]
\centering
\includegraphics[width=\textwidth]{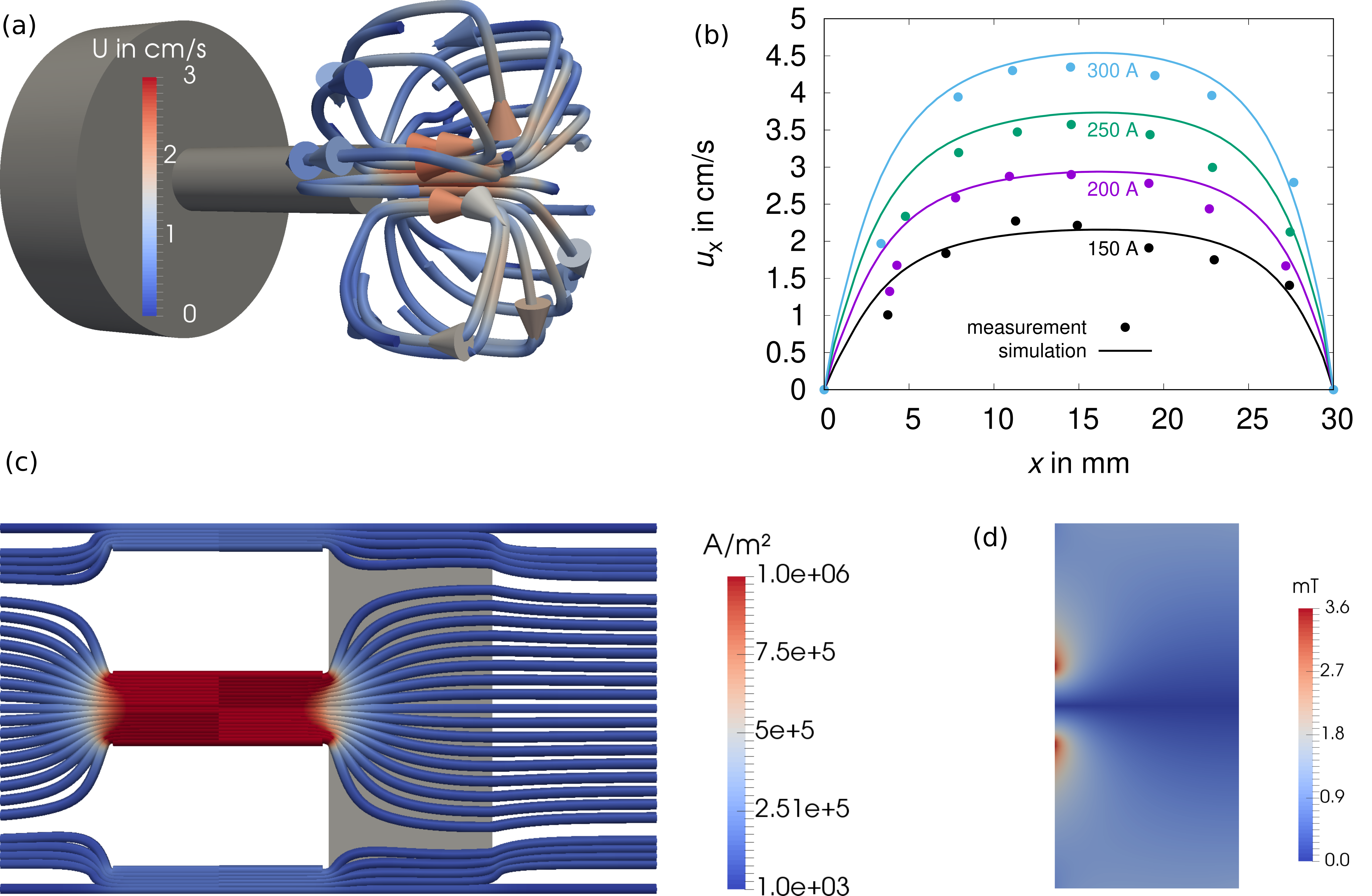}
\caption{Electro-vortex flow at $I=200$\,A (a) and measured axial
  velocity at the cylinder axis as given by Zhilin et
  al. \cite{Zhilin1986} (b) \nw{as well as the corresponding current
density (c) and the magnetic field (d).}}
\label{f:zhilin_flow}
\end{figure}

Figure \ref{f:zhilin_flow}a shows the general flow
structure \nw{(grid resolution in Hg 0.5\,mm)}. Assuming infinitely
long lateral current leads and
neglecting external magnetic fields, we expect exactly such a
symmetric flow. Further we expect the velocity along the cylinder axis
to be approximately uniform in the middle of the test section
(see fig. \ref{f:zhilin_flow}b) as long as the
current is not extremely low. The simulated curves for
$I=150\dots300$\,A fit very well to the measured velocity values
(dotted). A certain  deviation can be explained by the many unknown
experimental parameters; especially the length of the rod has a
certain influence on the magnitude of the flow \rw{(for details, see
\ref{s:detailsTestCase3}).}


\section{Summary and outlook}
We have developed a solver for electro-vortical flow, using a
mesh mapping method. Arbitrary solid and fluid conductors are fully
coupled. Electric potential and current density are solved on a global mesh,
and copied to the fluid mesh. This parent-child mesh technique is much
faster than the classical segregated approach. \nw{An improved
regularisation technique for the solution of the Poisson equation
of the electric potential is presented.}
The magnetic field is computed fully
parallelly using Biot-Savart's law. This was shown to be efficient
at least up to 64 processors. Calculating Biot-Savart's law only on the
boundaries and solving a corresponding induction equation in the fluid region
speeds up the magnetic field computation drastically. \nw{A first
   validation of the solver was done using the commercial
  software Opera and by comparison with experimental data.}

The solver presented  can easily cope with up to 1 million
cells. For larger simulations, a multigrid method or a coarser grid
for the magnetic field computation might be necessary. Further, the
solver shall be \nw{enhanced by a turbulence model and will be further
  validated using} recent experimental data. For a meaningful
comparision to experimental data, all dimensions of the setup and
all conductivities of the conductors as well the placement of the
feeding lines and possible magnetic background fields must be
known. Only in that case a computation of the experimentally investigated case
can successfully be performed. We aim to use the solver to study
electro-vortex flow in liquid metal batteries
\cite{Kim2013b,Weier2017,Ashour2017a} and aluminium reduction cells
\cite{Evans2007} as well as for related experiments
\cite{Starace2014, Kelley2014}. \nw{In the long term, a comparison
  between a segregated, a block matrice coupled (not existing in OpenFOAM yet) and
  the here presented electro-vortex flow solver is planned.}

\section*{Acknowledgements}
This work was supported by Helmholtz-Gemeinschaft Deutscher
Forschungs\-zentren (HGF) in frame of the Helmholtz Alliance
``Liquid metal technologies'' (LIMTECH). The
computations were performed on the Bull HPC-Cluster
``Taurus'' at the Center for Information Services and
High Performance Computing (ZIH) at TU Dresden and on the
cluster ``Hydra'' at Helmholtz-Zentrum Dresden - Rossendorf.
Fruitful discussions with R. Ashour, S. Beale, V. Bojarevics,
D. Kelley, A. Kharicha, J. Priede and F. Stefani on several aspects of
electro-vortex flow are gratefully acknowledged. N. Weber thanks
Henrik Schulz for the HPC support.

\section*{References}

\ifdefined\DeclarePrefChars\DeclarePrefChars{'’-}\else\fi

\appendix

\section{Detailed simulation of test case 3}\label{s:detailsTestCase3}
\begin{figure}[hbt!]
\centering
\includegraphics[width=0.7\textwidth]{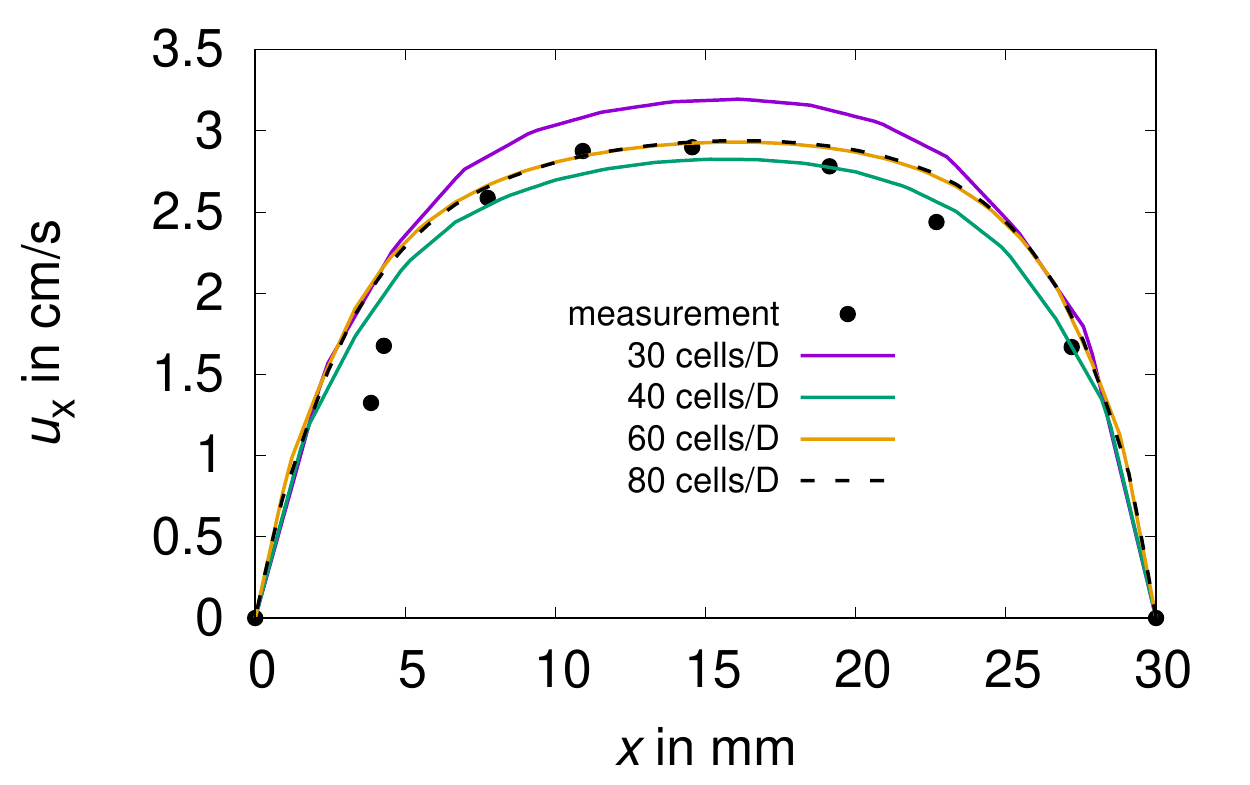}
\caption{Grid study for $I=200$\,A. The mesh is refined between 30 and
80 cells on the diameter; only cubic cells are used. The curves show
the steady state velocity along the jet.}
\label{f:grid}
\end{figure}

\begin{figure}[hbt!]
\centering
\includegraphics[width=0.7\textwidth]{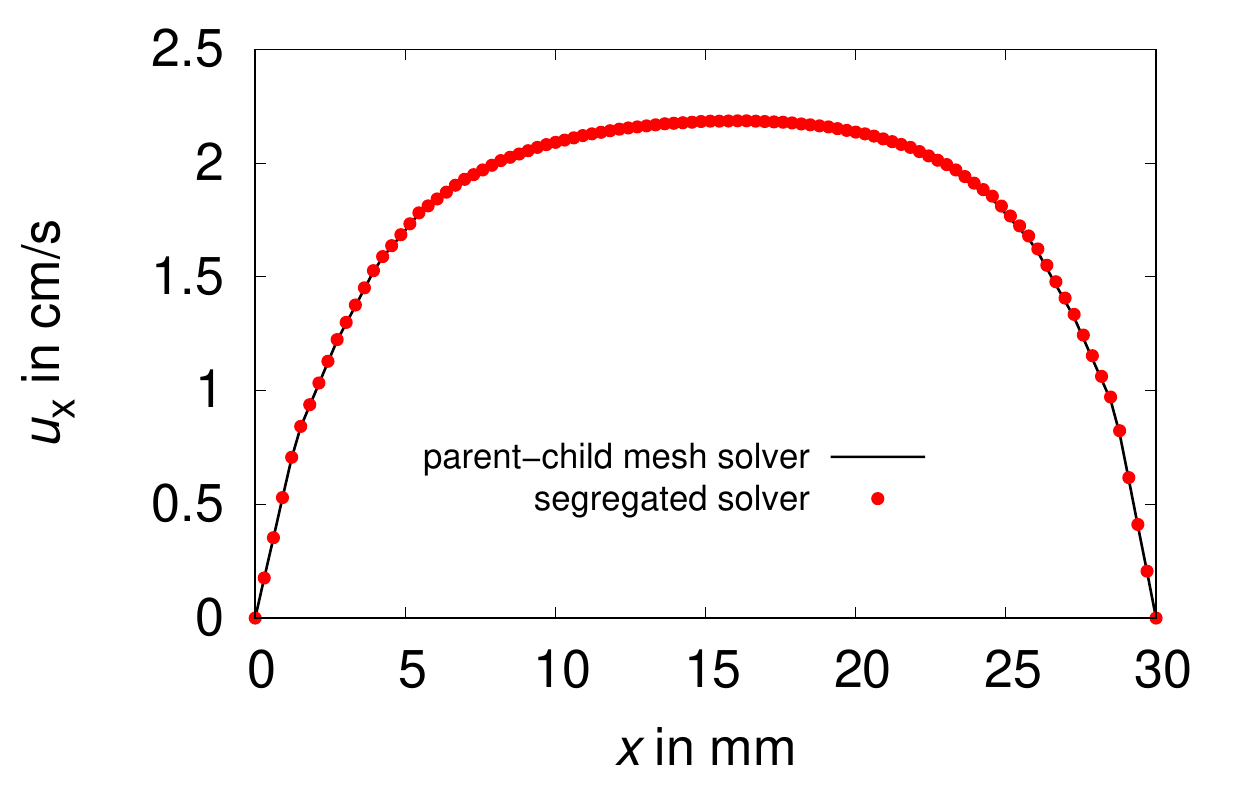}
\caption{Comparison of the segregated (see \cite{Weber2014b}) and
  parent-child mesh solver. The current is $I=200$\,A, 50
  cells/diameter are used.}
\label{f:cmpSegregated}
\end{figure}

\rw{
In this section we present further details of test case 3. Figure
\ref{f:grid} shows the grid study: at a current of 200\,A, 60 cells on
the diameter show convergence of the jet velocity. Figure
\ref{f:cmpSegregated} shows the steady state jet velocity at 200\,A
using an old segregated \cite{Weber2014b} and the here developed
solver. The results match perfectly; the coupled solver was more than
10 times faster than the segregated one.}

\begin{figure}[hbt!]
\centering
 \subfigure[]{\includegraphics[width=0.5\textwidth]{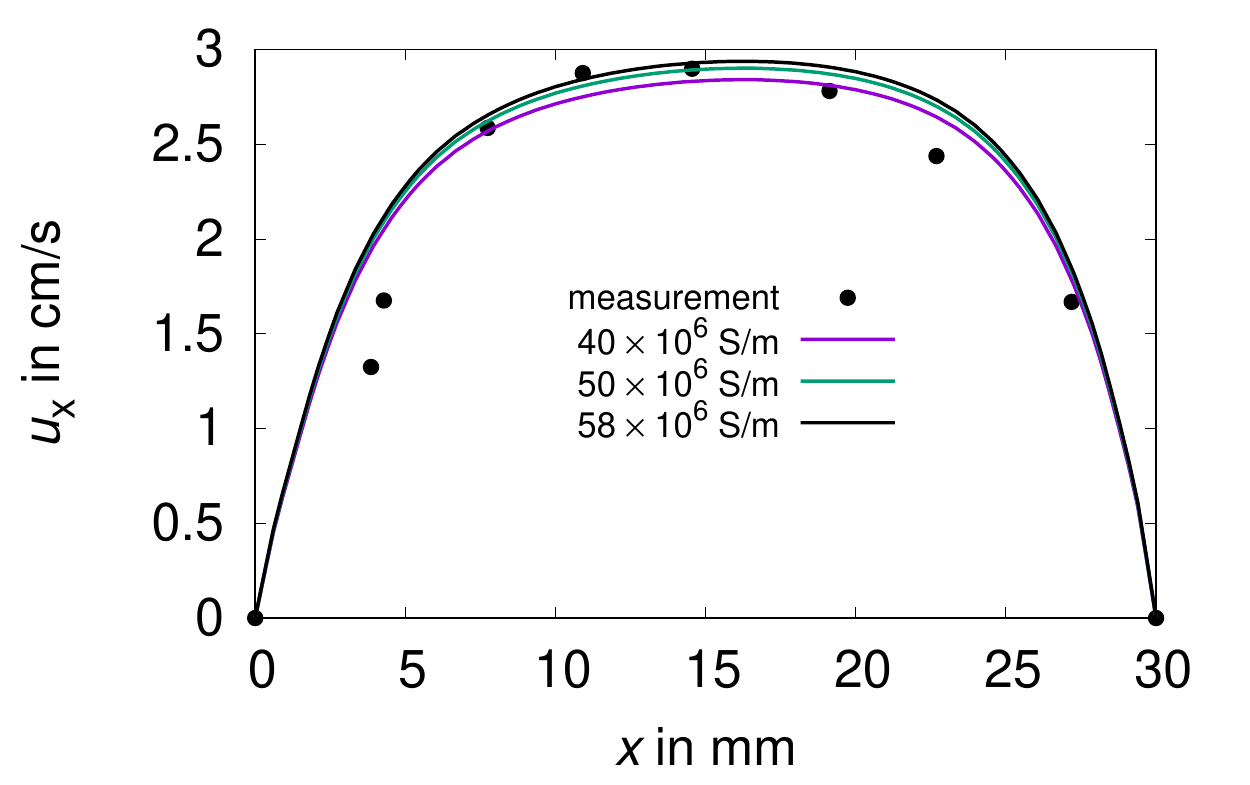}}\hfill
 \subfigure[]{\includegraphics[width=0.5\textwidth]{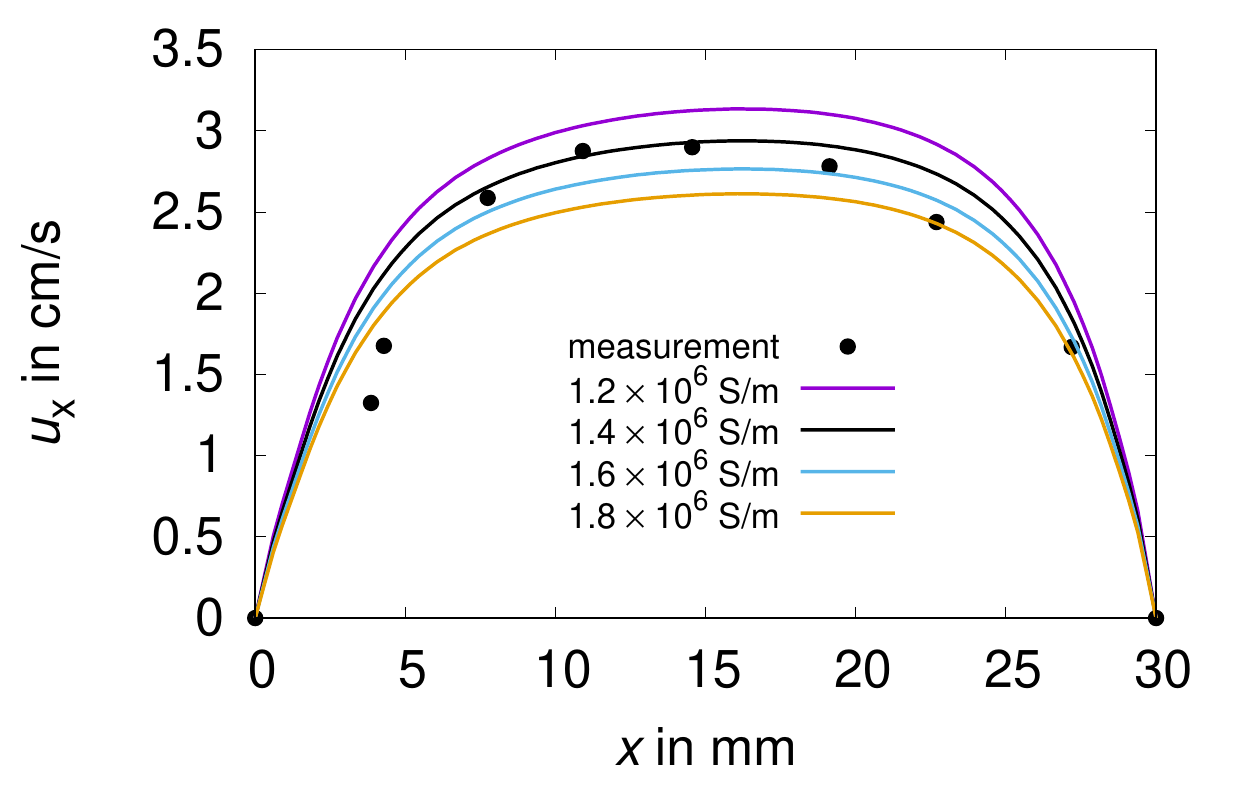}}\hfill
 \subfigure[]{\includegraphics[width=0.5\textwidth]{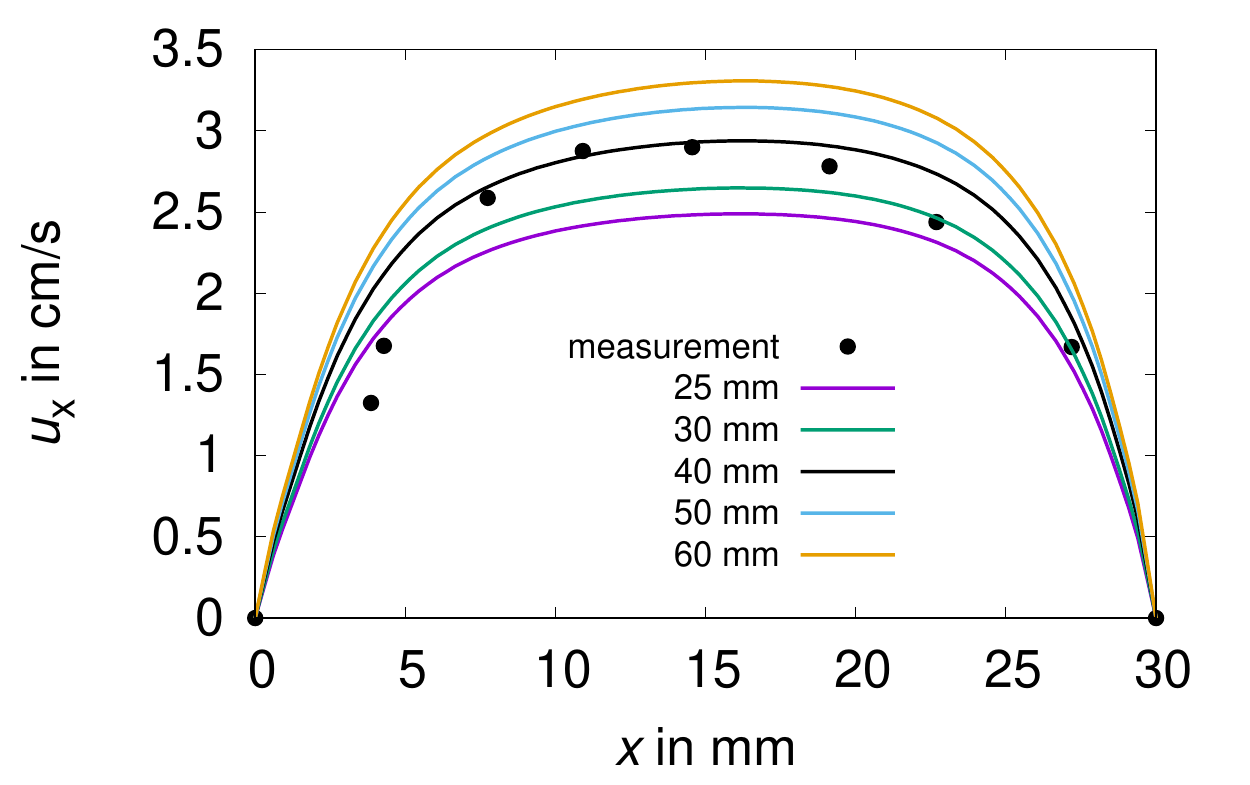}}\hfill
 \subfigure[]{\includegraphics[width=0.5\textwidth]{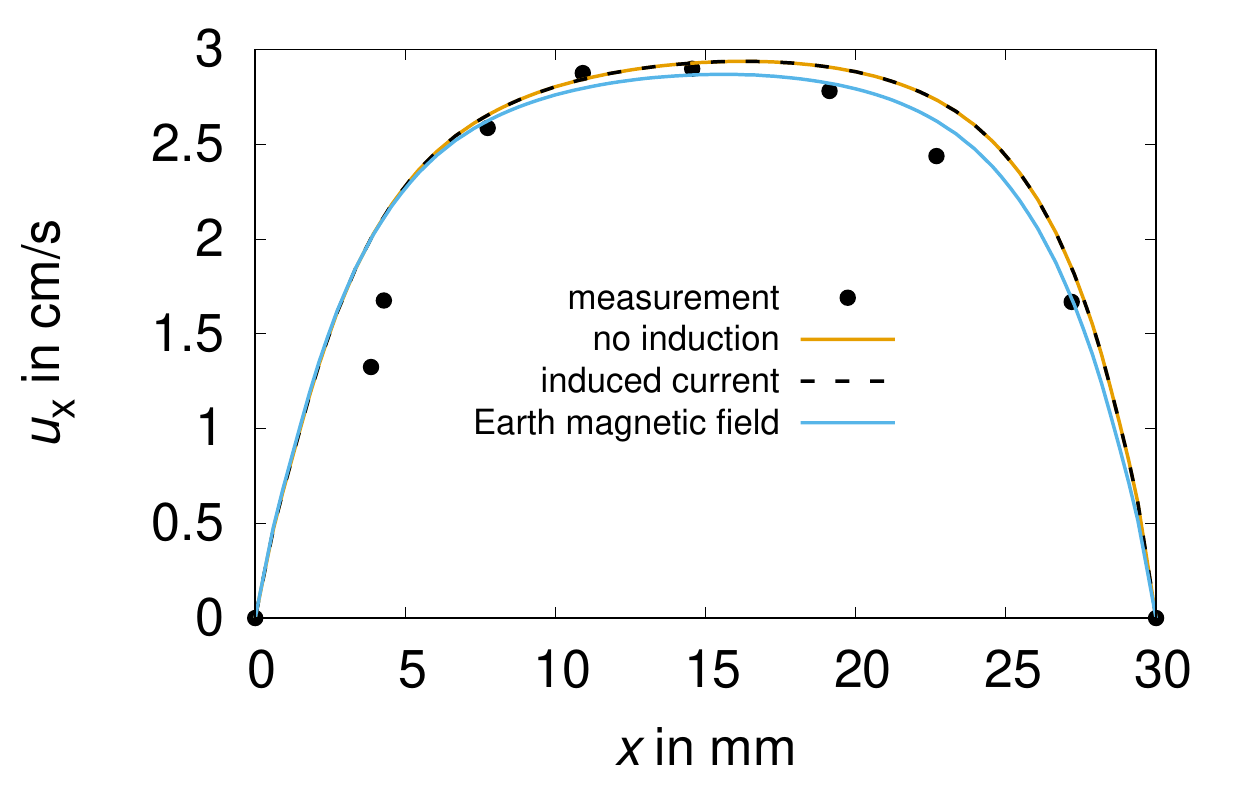}}
 \caption{Steady state flow velocity along the jet for $I=200$\,A. In
   (a) the conductivity of the copper electrodes is varied, in (b) the
   conductivity of the steel tube. A variation of the length of the
   rod (c) has a significant influence on the flow speed. Finally, (d)
   illustrates the influence of the Earth magnetic field and the induced
   current.}
\label{f:parameterstudy}
\end{figure}

\rw{
Finally, we illustrate in figure \ref{f:parameterstudy} the jet
velocity at 200\,A for a variation of the copper and steel electrical
conductivity and the length of the rod. These properties /
measurements are not exactly known from the experiment. While the
exact conductivity of the copper conductors is negligible, the
conductivity of the steel tube and the length of the rod change the
steady state velocity. We illustrate further the influence of the
induced current and the Earth magnetic field. We see, that the induced
current and magnetic field are negligible. However, the (vertical)
Earth magnetic field changes not only the speed of the jet, but also
its shape. We suspect, that other vertical magnetic fields (from the
feeding lines) may have an additional influence on the jet.}
\end{document}